# Surface modified sulfur nanoparticles can escape the glutathione reductase mediated detoxification system in fungi

Samrat Roy Choudhury[*], Arunava Goswami

*Biological Sciences Division, Indian Statistical Institute, 203 B.T Road, Kolkata 700108, India, [*] Phone: +919433337755, E-mail ID: samratroychoudhury@gmail.com*

**Abstract**

The antifungal effects of orthorhombic (~10 nm; spherical) and monoclinic (~50 nm; tetrapod) sulfur nanoparticles (SNPs) were studied against the NADPH-dependent glutathione reductase (GR) mediated xenobiotic detoxification system (GSH-GSSG) in filamentous fungi (*Aspergillus niger* as a model organism). Both the SNPs induced significant reduction in fungal growth and spore formation, and also introduced marked deformities on the surface of conidiophores at their sub-inhibitory concentrations. A genome wide transcriptome profile then revealed manifold reduction in the expression of GSH-GSSG transcripts among SNPs treated fungal isolates, which is unusual for the micron sized elemental sulfur but probably effective in terms of antifungal efficacy.

**Introduction**

The purifying and beneficial properties of Elemental Sulfur (ES) are being explored since time immemorial (Mitchel 1996). It is interesting to note that while ES plays a crucial role in protein synthesis at a basal quantity, a high concentration of ES is however, considered to be toxic to microorganisms (Owens 1963). Nanosized dimension and suitable surface modifications of ES further augments its antimicrobial efficacy. In our previous studies, the antifungal properties of orthorhombic (α) and monoclinic (ß-) allotropes of sulfur nanoparticles (SNPs) have already been discussed (Roy Choudhury et al. 2011; Roy Choudhury et al. 2013a). Briefly, in the microbial cytosol both α- and ß-SNPs generate free reactive thiols (-RSH), which induce ROS-stress, impairs a cluster of respiratory enzymes, and/ or denatures certain proteins and lipids (Roy Choudhury et al. 2012). SNPs also enter into the mitochondrial matrix and influence fungi to undertake the alternative glyoxylate shunt in favor of energy conservation (Fig.1) (Roy Choudhury et al. 2013b).

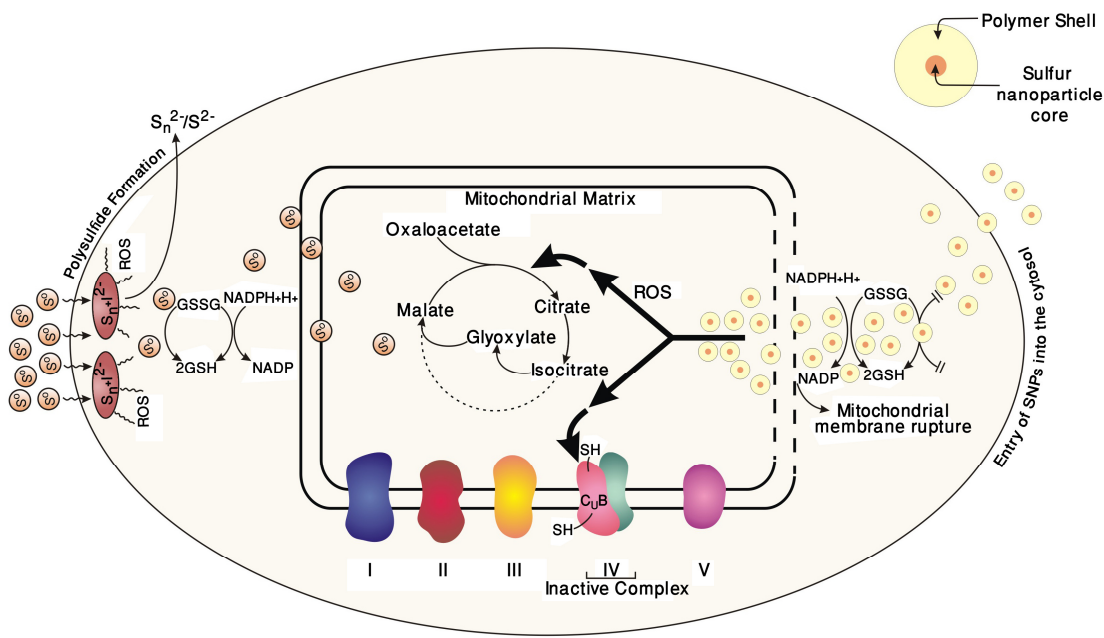

*Fig.1 A schematic cartoon depicts the comparative route of micronized elemental sulfur (ES; S0), and sulfur nanoparticles (SNPs) into the mitochondrial matrix of fungi. ES forms long polysulfide chain upon entering into the microbial cytosol and readily removed as sulfides ($S^{2-}$ /$S_n^{-}$) by the NADPH-dependent glutathione (GR) mediated detoxification system in fungi. In contrast, polymer coated SNPs surpass the GR system in fungi, disrupt mitochondrial membranes, and inhibit cytochrome-c-oxidase (complex-IV) of the oxidative phosphorylation.*

In general, GSH synthetases (γGCS and GS), NADPH dependent GSH-regenerating reductase (GR), glutathione transferase (GST) along with peroxide eliminating glutathione peroxidase ($GP_X$) and glutaredoxins (Grxs/Trx) are directly involved in the elimination oxidative compounds in yeasts and other fungi (Costa and Moradas- Ferreira 2000; Moradas-Ferreira and Costa 2001). However, no reports to the date are available essaying the effects of SNPs against the GR mediated antioxidant system in fungi. Herein, we report a non-targeted cDNA microarray based genome-wide transcription analysis to determine the oxidative effects of SNPs against the NADPH dependent GR-mediated detoxification system in *Aspergillus niger*. The disturbing effects of SNPs against the fungal growth, morphology, and sporulation (Fig.2) were then correlated to the alteration in transcript expression of the GSH metabolism components. A simultaneous study was also undertaken to determine the effect of SNPs treatment on the GR enzymatic activity.

**Material and Methods**

**Preparation and characterization of sulfur nanoparticles**
Polyethylene Glycol encapsulated (PEGylated) α-SNPs were prepared via liquid phase precipitation method (Roy Choudhury et al. 2011). Span-80 (sorbitan monooleate)-Tween-80 (polyethylene glycol-sorbitan monooleate) encapsulated β-SNPs were prepared via water-in-oil microemulsion technique (Roy Choudhury et al. 2013a). SNPs were characterized for their physicochemical properties as described in our previous studies (Roy Choudhury et al. 2013a; Roy Choudhury et al 2013b). High resolution-transmission electron microscopy (HR-TEM) (2010F, JEOL Ltd., Tokyo, Japan) was carried out at 200 kV on a carbon-coated copper grid to confirm the actual size of SNPs. Stock concentration of the prepared nanocolloids was evaluated as per our previously published method (Kumar et al. 2011).

**Fungal strains and growth conditions**
The wild strain of *A. niger* was isolated from rotten potato tubers, obtained from the commercial sources in Kolkata, India. Tubers with visible lesions of black moulds were then disinfected with 1% sodium hypochlorite solution for 3 minutes, followed by thorough washing with Millipore water (Sartorius Stedium biotech, arium 611VF and arium 61316 systems, Aubgene, France). The mycelium of the isolated fungal strains were placed in 5 ml of water and subjected to vortex (Genei Pvt. Ltd, Bangalore, India). The content was then poured into petridishes containing saboraud dextrose agar (SDA; Himedia laboratories Pvt. Ltd., Mumbai, India) for rapid and viable fungal growth. The petri-dishes were incubated at a temperature of 28°C for 48 hrs. The germinated conidia were then transferred to new petri-dishes containing SDA and subcultured regularly at an interval of 10 days. Simultaneously, the conidia were transferred into 10 % sterile glycerol and stored at - 80 °C for long term use. Species specific studies with partial gene sequences of ITS / 5.8 rRNA (NCBI accession number HQ293217) and β-tubulin (NCBI accession number HQ293218) were done prior to their deposition in Microbial Type Culture Collection (MTCC); India (accession number 10180). MTCC-282 and MTCC-2196 strains were obtained from, MTCC, IMTECH-CSIR, Chandigarh, Govt. of India.

**The Antifungal assays with SNPs**

The antifungal effects of SNPs were determined in terms of reduction in the radial growth of fungi using a modified agar dilution method (ADM). The microbial strains were grown in potato dextrose agar (PDA) and treated with serially diluted (two fold) concentrations of SNPs (Roy Choudhury et al. 2013b) to determine minimal-inhibitory concentrations (MICs) and sub-inhibitory concentrations (SICs). Inhibitory effects of SNPs on sporulation of the fungi were determined with a slide bioassay (Resende et al. 1996; Roy Choudhury et al. 2011) Deformities at the surfaces of treated fungal hyphae were visualized with Field Emission Scanning Electron Microscopy (FE-SEM) [JEOL JSM-600F, Tokyo, Japan] at 5.0 kV vacuum under x 2500 magnification.

**Microarray analysis for the GR transcripts**

Expression of the crucial transcripts of GSH metabolism and TRX/GRX pathway were studied with microarray analyses in triplicate. Total mRNA (with RNeasy minikit, Qiagen, CA, USA) from the untreated, ES and SNPs-treated (at their sub-inhibitory concentrations) of strain MTCC-10180 was extracted to determine the induction/ repression level. Gene expression of the purified mRNAs were determined on an 8 X 15 K array slide (AMADID: 033147, Agilent Technologies Inc., CA, USA) using one color microarray based gene expression analysis kit (Agilent Technologies Inc., CA, USA). Individual gene hybridization intensities were normalized to balance their uniform expression using GeneSpring GX-11.5 software (Agilent Technologies Inc., CA, USA). Fold induction or repression in gene expression was calculated in a $\log_2$ scale. A negative or positive $\log_2$ value indicated a reduced or induced transcript level, respectively. Spurious or anomalous results, which had been flagged by the ScanAlyze (http://rana.lbl.gov/EisenSoftware.htm) program, were deleted from the data set. Gene ontology, molecular functions, and tracing of specific genetic pathways were identified, based on *Aspergillus niger* v 3.0 genome project (http://genome.jgi-psf.org/Aspni5/Aspni5.home.html).

**Measurement of enzyme activity of GR**

Fungal strains were incubated with SNPs supplemented media containing (IAA) iodoacetic acid. IAA significantly deplete the GSH content of cells (Schmidt and Dringen 2009), and hence impair the GR-mediated antioxidant efficacy. The SICs to IAA were fixed depending upon the tolerance limit of microbes to IAA. Changes in intracellular GR-enzyme activity was then measured from the SNP treated isolate of *A. niger* (MTCC-10180), both in presence and absence of IAA. GR activity was measured following the originally proposed method (Carlberg et al. 1985). Briefly, Around 0.2 gm of liquid nitrogen frozen mycelial mass (scraped from PDA plates) was extracted in the extraction buffer containing 50 mM Tris-HCl (pH: 7.8), 0.1 mM EDTA, 0.2% TritonX-100, 1mM PMSF, and 2 mM DTT. The homogenate was centrifuged at 14000 rpm for 30 min at 4 ºC. The supernatant (enzyme sample) was then mixed with 25 mM Na-PO4 buffer, pH 7.8, 5mM GSSG and 1.2 mM NADPH.Na4 assay solutions. For 3 ml of total reaction volume, 25 mM Na-PO4 buffer, 5 mM GSSG, 1.2 mM NADPH.Na4, sample were added in a ratio of 7:1:1:1. The oxidation of NADPH by GR was estimated from the decreased absorbance at 390 nm, recorded continuously for 180 seconds. The GR activity was calculated as Unit/mg protein/min per sample, using the following formula:

C0 (original concentration) =    Absorbance x dilution factor/

$\varepsilon$ (extinction coefficient of NADPH 6.2 mM-1cm-1 at 390 nm)

**Results and discussion**

**Physicochemical and antimicrobial properties of SNPs**

Both the SNPs were prepared in colloidal form. TEM micrographs revealed that α-SNPs were nearly spherical in shape with an average particle size (APS) of ~10 nm (Fig.2A). β-SNPs looked like tetrapod with APS of ~50 nm (Fig.2B). MIC of α- and β-SNPs was evaluated to be 8000 ppm and 32000 ppm respectively. ADM showed gradual reduction in the radial growth of fungi along the increasing concentration gradient of SNPs (Fig.3; a-c) (Roy Choudhury et al. 2011).

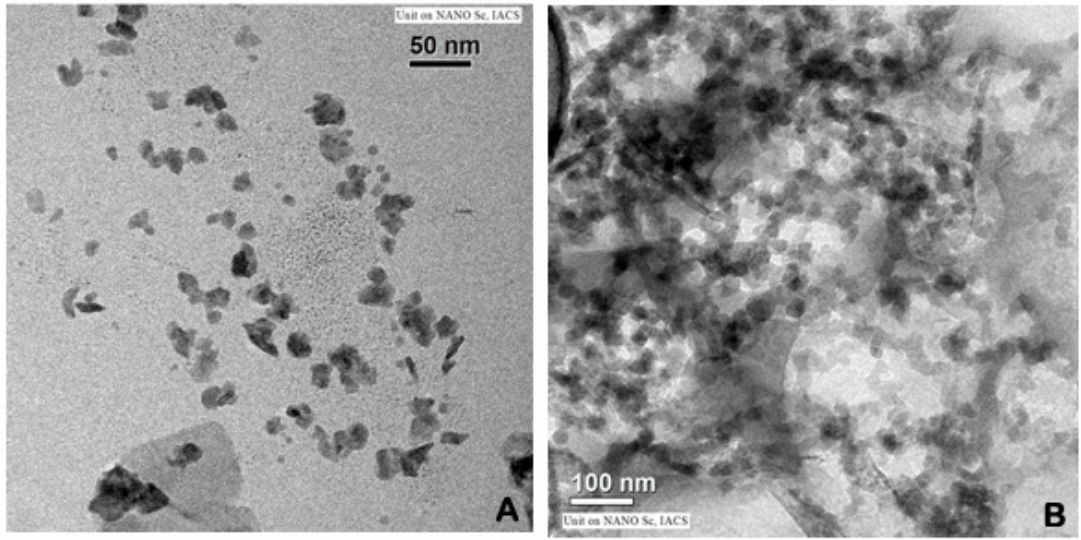

*Fig.2* TEM micrographs of α-SNPs (average Particle Size ~10 nm; spherical) and β-SNPs (APS~50 nm; tetrapod)

At higher doses, SNPs also found to obliterate the rate of sporulation (Fig.3; g-i) and introduce significant deformities to the surface of conidiophores (Fig.3; d-f) (Roy Choudhury et al. 2013b). In contrast, ES failed to exert any marked antifungal effects.

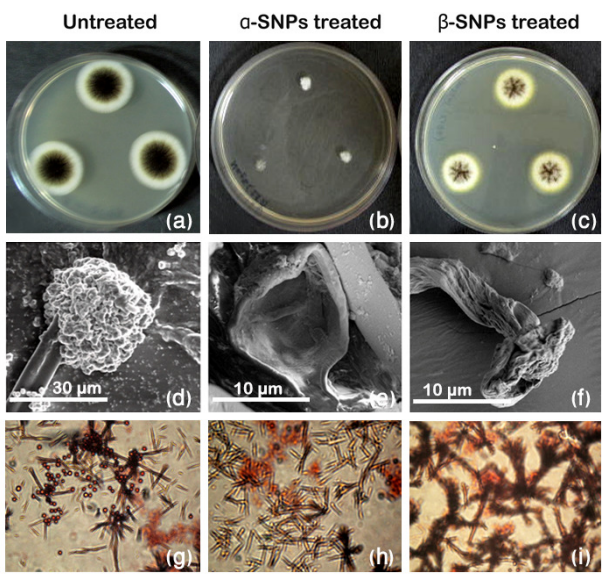

*Fig.3* α- and ß-SNPs inhibit radial growth (a-c), and spore count (g-h) of *Aspergillus niger*. SNPs also induce morphological anomalies of the conidiophores (d-f) at their sub-inhibitory concentrations. Fig.2; d-f were reproduced from Roy Choudhury et al. (2013b).

**Alteration in the GR-transcript expression**

SNPs mediated antifungal phenotypes were extrapolated to their interfering effects of the fungal antioxidant system. Filamentous fungi are equipped to use distinct antioxidant enzymes of the GSH-GSSG pathway to reduce the load of various form sulfur toxicity. For example, NADPH dependent GR readily converts ES to soluble polysulfide, and makes their penetration difficult against microbial membranes (Sato et al. 2011). A lesser fraction of micronized ES, if succeeds to enter the microbial cytosol to generate toxic $H_2S$; GSH transporter γ-L-glutamyl-cysteine synthase (GCS) promptly acts to scavenge it from the cell. Some of the sulfurous xenobiotics have potential to react either spontaneously with the –SH group of GSH or to form GSH S-conjugates (GS-X), which then become susceptible to nucleophilic attack by GSH on the electrophilic sulfur by the aid of glutathione S-treansferases (GSTs) (Pócsi et al. 2004). Transcriptome data of the present study revealed that ES treatment differentially induced expression of the GSH-GSSG detoxification transcripts, as expected (Fig.4).

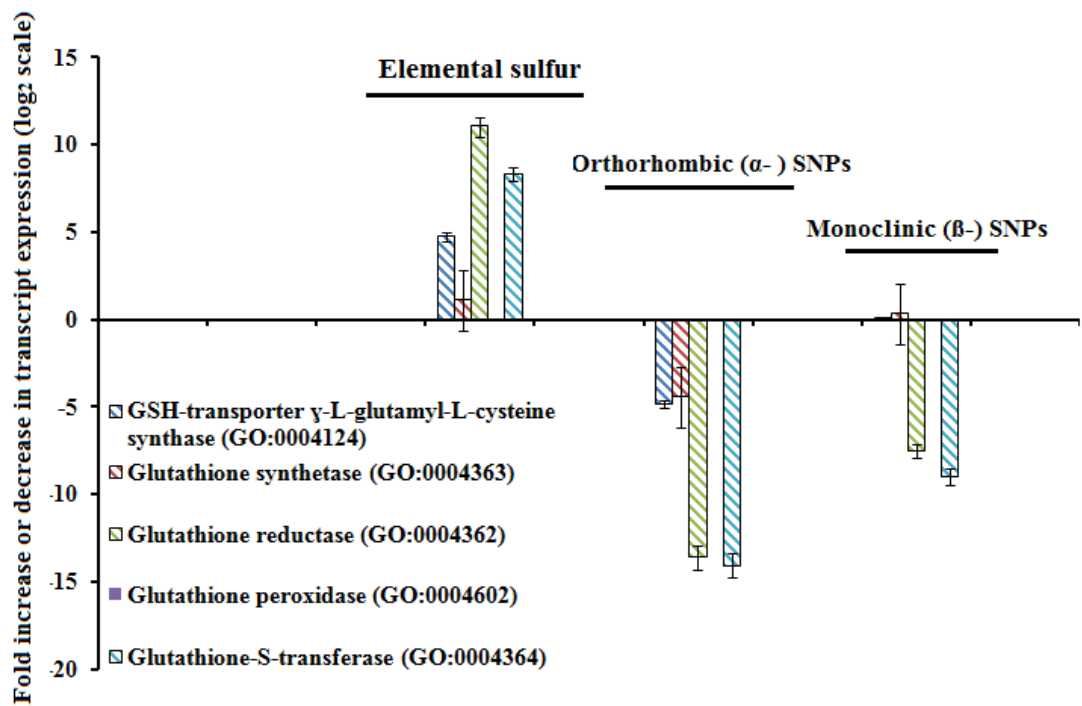

*Fig.4 Induction or repression in the transcript expression of GSH-GSSG pathway components after treatment with micronized elemental sulfur, α- and β-SNPs.*

In particular, the highest fold of induction (11.02) was observed for GR, which also suggests the occurrence of proactive NADPH mediated GR system in the tested *A. niger* strain. GCS and Glutathione synthetase of the GSH-GSSG pathway were also reasonably elevated. Noteworthy, a substantial elevation (8.312) in GST expression was also noticed after ES treatment, which indicates that ES may provoke the fungal antioxidant system in more than one way. In contrast, transcript expression for each of the GSH-GSSG components was repressed manifold followed by the α-SNPs treatment. For ß-SNPs, though the GCS and glutathione synthetase increased over 2 fold, GR and GST expressions were found to be repressed. It's worth mentioning that GPx expression level was not changed after ES/SNPs treatment, which may correlated to its non-responsiveness to the sulfur induced stress. Moreover, components of TRX/GRX were found to remain compromised/ not detected after ES treatment, however repressed differentially with SNPs treatment (Table-1).

*Table 1: Change in transcript profile of the thioredoxin/glutaredoxin (TRX/GRX) pathway in A. niger after treated with elemental sulfur, orthorhombic, and monoclinic sulfur nanoparticles (SNPs)*

| **TRX/GRX Pathway** | Elemental Sulfur | Orthorhombic SNPs | Monoclinic SNPs |
|---|---|---|---|
| Thioredoxin (GO:0019379) | Not detected | -5.13 ± 2.89 | -3.23 ± 1.38 |
| Glutaredoxin, glutathione dependent oxidoreductase (GO:0015035) | Not detected | -10.06 ± 3.12 | -9.75 ± 1.42 |
| Thioredoxin reductase (GO:0004791) | Not detected | -8.8 ± 1.88 | -9.5 ± 4.02 |
| PAPS reductase (GO:0004604) | Not detected | -5.13 ± 2.02 | -1.02 ± 1.11 |
| Thioredoxin peroxidases (GO:0009031) | Not detected | -10.83 ± 0.99 | -3.92 ± 2.01 |

The disparity in transcript expression followed by ES and SNPs suggest their completely different mode of penetration, and effect against the fungal antioxidant mechanism. In our previous studies we observed that SNPs can introduce various degrees of rupture in the mitochondrial membranes and impair Cytochrome-C-oxidase of oxidative phosphorylation (Roy Choudhury et al. 2013b).

**Alteration in the GR-enzyme activity**

GR-enzyme activity was found to be altered marginally with the increasing concentration gradient of α- and β-SNPs., but elevated at the SICs of SNPs (Fig.5).

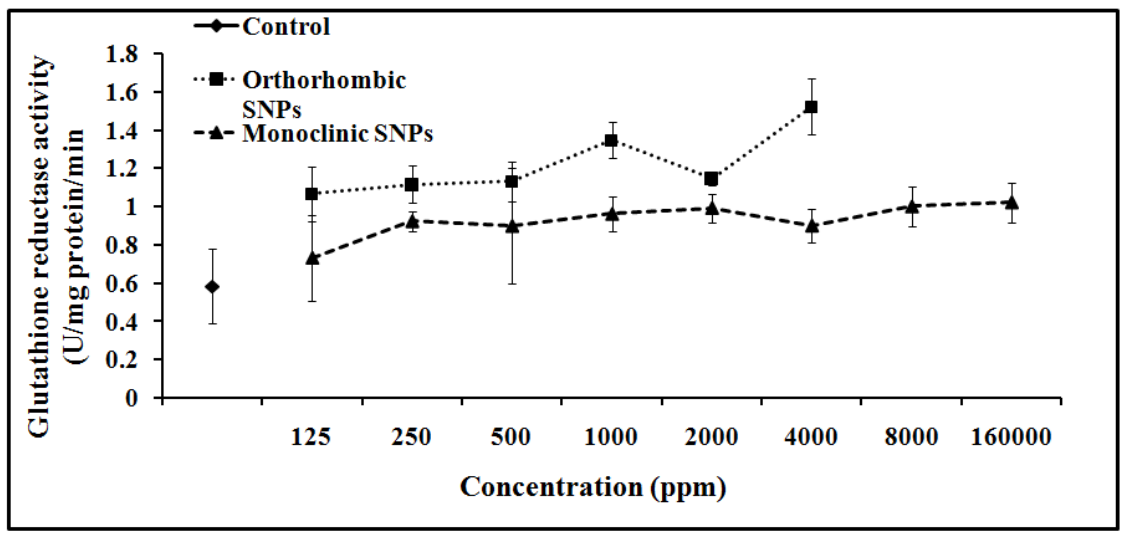

***Fig.5*** *Effect of α- and β-SNPs on the change of GR activity of A. niger. Averages and standard deviations were calculated from three independent rounds of experiment. (Reproduced from Roy Choudhury et al. 2013b)*

However, no GR activity was recorded in presence of IAA, and hence, the MICs of both α- and β-SNPs dropped down significantly (Roy Choudhury et al. 2013b).

**Discussion**

The present study additionally suggests that diminutive size (nano) and polymer stabilized surface of SNPs may camouflage their presence and facilitate their entry to the mitochondrial matrix by surpassing the proactive GSH-GSSG mediated detoxification systems of fungi. Moreover, the antifungal assay used here employs ADM, where the fungus has unavoidable access of ES and SNPs along the course of its development. Hence, even an abrupt increase in GR enzyme activity was observed at the SICs of SNPs; this post-translational elevation in GR activity turns out to be inefficient for SNPs-detoxification. In conclusion, for the first time we have explored the interaction of SNPs against the GR mediated detoxification system in fungi at transcriptome level. Nano-size and novel mode of action of SNPs make them suitable as next generation antifungal agent when ES fails, and fungi will probably take longer to find ways to resist them.